# Imaging and tuning polarity at SrTiO$_3$ domain walls

Yiftach Frenkel[1], Noam Haham[1], Yishai Shperber[1], Christopher Bell[2], Yanwu Xie[3,4,5], Zhuoyu Chen[5], Yasuyuki Hikita[3], Harold Y. Hwang[3,5], Ekhard K.H. Salje[6,7] and Beena Kalisky[1,*]

1. Department of Physics and Institute of Nanotechnology and Advanced Materials, Bar-Ilan University, Ramat-Gan, 5290002, Israel
2. H. H. Wills Physics Laboratory, University of Bristol, Tyndall Avenue, Bristol, BS8 1TL, UK
3. Stanford Institute for Materials and Energy Sciences, SLAC National Accelerator Laboratory, Menlo Park, California 94025, USA
4. Department of Physics, Zhejiang University, Hangzhou, 310027, China
5. Department of Applied Physics, Geballe Laboratory for Advanced Materials, Stanford University
476 Lomita Mall, Stanford University, Stanford, California, 94305, USA
6. Department of Earth Sciences, University of Cambridge, Downing Street, Cambridge CB2 3EQ, United Kingdom
7. State Key Lab Mech. Behav. Mat., Xi'an Jiaotong University, Xi'an 710049, China
* beena@biu.ac.il

**Electrostatic fields tune the ground state of interfaces between complex oxide materials. Electronic properties, such as conductivity and superconductivity, can be tuned and then used to create and control circuit elements and gate-defined devices. Here we show that naturally occurring twin boundaries, with properties that are different than their surrounding bulk, can tune the LaAlO$_3$/SrTiO$_3$ interface 2DEG at the nanoscale. In particular, SrTiO$_3$ domain boundaries have the unusual distinction of remaining highly mobile down to low temperatures, and were recently suggested to be polar. Here we apply localized pressure to an individual SrTiO$_3$ twin boundary and detect a change in LaAlO$_3$/SrTiO$_3$ interface current distribution. Our data directly confirm the existence of polarity at the twin boundaries, and demonstrate that they can serve as effective tunable gates. As the location of SrTiO$_3$ domain walls can be controlled using external field stimuli, our findings suggest a novel approach to manipulate SrTiO$_3$-based devices on the nanoscale.**

The interface between Strontium titanate (SrTiO$_3$, STO) and Lanthanum Aluminate (LaAiO$_3$, LAO) hosts a gate tunable two dimensional electron gas (2DEG)[1–4]. It has been demonstrated that the 2DEG can be confined to create devices such as an interfacial gate defined SQUID[5] or a single electron transistor[6,7]. Here we show that the rich physics of STO can be utilized to control the conduction at the interface.
STO undergoes a ferroelastic phase transition at 105 K. In the ferroelastic phase the material forms a dense network of twin domains with well-defined boundaries between each set of twins[8]. At ~37 K STO goes through another anomaly where its dielectric constant starts to diverge but a macroscopic ferroelectric state is suppressed by zero point fluctuations leading to a quantum paraelectric state[9]. Additional symmetry breaking at low temperatures originates from Sr ions moving along the [111] direction resulting in triclinic symmetry [10]. More recently Scott *et al.*[11] showed by Resonant Ultrasonic Spectroscopy that domain walls in STO can indeed be polar. Salje *et al.*[12] confirmed this finding by piezoelectric spectroscopy measurements and detected that weak polarity resides widely in STO below ~80 K and becomes strong below ~40 K. They concluded that polarity is generated 'only on the nanoscale and not as a bulk homogeneous property'. As complex domain and domain wall structures are widespread in STO, polarity was found to encompass large parts of the STO sample but still emanated from domain walls (see suppl. material in ref 12). Polarity at the domain walls[12] was argued to be similar to that of CaTiO$_3$ due to displacement of the Ti atoms inside un-tilted oxygen octahedra inside the domain walls[13,14]. In the presence of such polarity the obvious question that immediately arises is how the domain walls affect the electronic properties of nearby conducting layers.

In 2009 Seidel *et al.* observed conductance along ferroelectric domain walls of the insulating multiferroic $BiFeO_3$ (BFO),[15] setting the ground for an enormous amount of work in the field. Later studies by Whyte *et al.*[16] and Crassous *et al.*[17] showed significant advances in the creation and control of the domain walls taking the field another step towards realization in real devices. We consider a different case, namely polar domain walls in non-polar STO and their effect on interfaces with $LaAlO_3$ (LAO).

The conducting interface formed between LAO and STO[18], provides the opportunity to examine the influence of STO twin walls on the two dimensional conducting layer[19]. Local scanning probe mapping of the current flow[20] and electrostatic charge[21] as well as low temperature scanning electron microscopy (LTSEM)[22] of the LAO/STO interface revealed that the electronic properties are indeed modulated over STO domain walls. Recently Ma et al. suggested that ferroelectricity at the walls is induced above a threshold of applied electric field[22]. Here we image the wall polarity below 40 K and show that the walls are *intrinsically polar*. We suggest that this polarity is the mechanism responsible for the previously reported modulated current flow at the LAO/STO interface[20]. This is supported by the appearance of spatial modulations in the current flow only below 40 K, the onset temperature of strong wall polarity[12].

Stress is known to control both the domain wall polarity and the dielectric properties of STO[23,24]. In this work we examined the effect of stress on local electronic properties and the manner in which it controls the properties of the overlaying LAO/STO interface. By focusing on individual domain walls, we provide direct observation of polarity at the walls. We suggest that the local stress tuned the wall polarity, thus depleting or accumulating electrons near the wall. We examine the prospect of using the walls as local electrostatic gates.

In order to investigate individual boundaries, we constructed a scanning stress microscope, to map the electrical response to local stress as a function of lateral position (Methods). We found that a relatively small *local* stress induces a strikingly strong influence on the *global* conductance of the device. Mapping the response over the sample clearly identified the domain walls as the electrically active locations. A dramatic aspect of this effect is that the influence of the local stress propagates relatively long distances from the point of application.

The scanning stress microscope consists of a non-conducting silicon tip which was rastered over a conducting LAO/STO interface patterned in a square van der Pauw geometry of 200 μm x 200 μm. Piezoelectric elements were used to push the tip into contact with the sample and apply local stress with a contact area of $(0.1-1~\mu m)^2$, (Methods). The tip applied forces up to 2 μN, exerting stress gradients that decay as the square root of the distance from the contact point. The macroscopic four probe voltage of the device was hence mapped as a function of the location of the tip for a specific stress at each point (Fig. 1a). When the tip pressed on the sample at specific locations the transport value of the whole 200 μm x 200 μm square changed significantly. A map of the change in the global device resistance versus the location of the applied stress could then be created (Fig. 1b). We confirmed that when the tip was scanned above the sample without contact, or the stress was applied outside the 200 μm x 200 μm square, the resistance of the sample did not change, as expected.

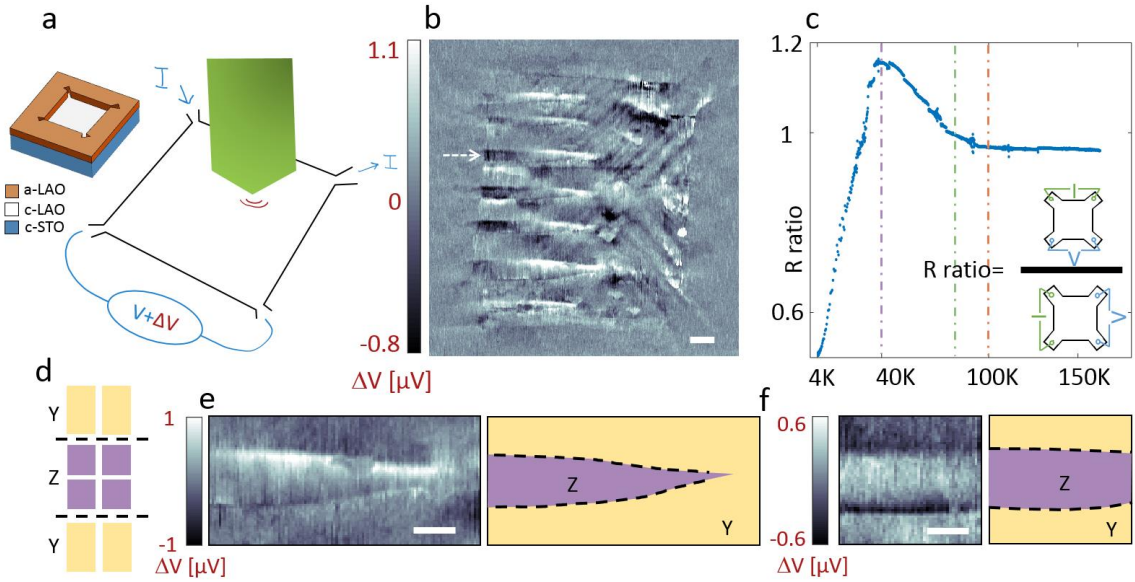

*Figure 1: Strong response to stress on STO domain walls.* (a) Illustration of the device and the experiment. The prefix "a" stands for amorphous, and "c" for crystalline. A non-conducting silicon tip is brought in contact with the sample. Scanning and vertical stress application was performed using piezo elements. Voltage measured in opposing leads (bottom pair) to current injection (top pair) detecting the voltage change ΔV. (b) ΔV as a function of location of the contact point reveals strong responses on domain walls. In this case the tip stress was 0.4 μN uniformly over the image. The background voltage V is 925 μV at 4.2 K. White arrow points to a needle [100] domain on which ΔV changes sign. Scale bar is 20 μm. (c) Transport data taken during cooldown. The ratio between the voltage measured in two perpendicular measurement orientations shows three meaningful temperatures: 105 K (red dashed line, breaking of unit cell symmetry), 80K (green dashed line, onset of weak domain wall polarity[12]) and 40 K (purple dashed line, strong polarity at the domain walls[12]). (d) Illustration, top view of [001] twin boundary between domains with unit cells elongated along the [010] (Y) and [001] (Z) original cubic crystallographic directions. (e-f) ΔV on two needle domains and illustration of the walls. ΔV signal clearly peaks at the walls. In the illustration the identity of the domains was chosen arbitrarily between Y and Z. The [100] boundary is always between Y and Z domains, as indicated by the dashed line in (d); a detailed description of the domain wall direction is shown in the Supplementary Information in Fig. S1. Scale bar is 10 μm.

The *ΔV* map shows stripe patterns in the [100] [110] [$\bar{1}$10] STO crystallographic directions (Fig. 1b). The sharp features in the map are 0.5 μm wide, which is the spatial resolution of our pressure tip, determined by the contact area and the shape of the tip. We identify these stripes as STO domain walls, based on the following: (a) the stripes are orientated along STO crystallographic directions; (b) the stripes configuration changes after cycling the temperature around the structural transition at 105 K (see Supplementary Information, Fig. S3); (c) we compare *ΔV* map with maps of the current flow obtained by scanning superconducting quantum interference device (SQUID). The configuration of the modulated current streaks over STO domains[20,25,26] is similar to the *ΔV* map recorded simultaneously (Supplementary Information Fig. S4).

Generally in STO, the domains can be structured on scales (down to tens of nanometres[27,28]) much smaller than our resolution. In order to analyse individual twin walls, we make use of twin needle domains with two well separated walls. The needle shape allows us to investigate one type of domain penetrating into the other as illustrated in Figs. 1e and f. In this manner we can observe a single wall. Scanning over a single domain wall shows that ΔV peaks at the wall. This indicates a change in the local current flow when the stress is applied to the domain wall. We note that in addition to the signal detected on domain walls, we also observe a relatively weaker resistance contrast when applying stress inside the needle domain. This could be related to the size and shape of our stress kernel, and can be further examined using a sharper tip. Here we focus on the more dominant effect at the domain walls. The dashed arrow in Fig. 1b points to another interesting feature, a change in the sign of $\Delta V$ along the domain wall.

In order to verify that the $\Delta V$ signal is stress induced, and is not generated by other magnetic or electric influences of our scanning system, we mapped $\Delta V$ at different locations in and out of the sample and as a function of height above the sample. We detected $\Delta V$ only after we made contact with the sample (Fig. 2a and b). After contact $\Delta V$ increased linearly with the applied stress, confirming that its origin is stress. We found that the response scaled with the domain size (Fig. 2c). The temperature evolution of the $\Delta V$ signal (Fig. 2d) is consistent with the enhancement in wall polarity below 40 K[12] and the temperature dependence of the interfacial current modulations[20]. This behaviour is key for understanding our data and for identifying its origin.

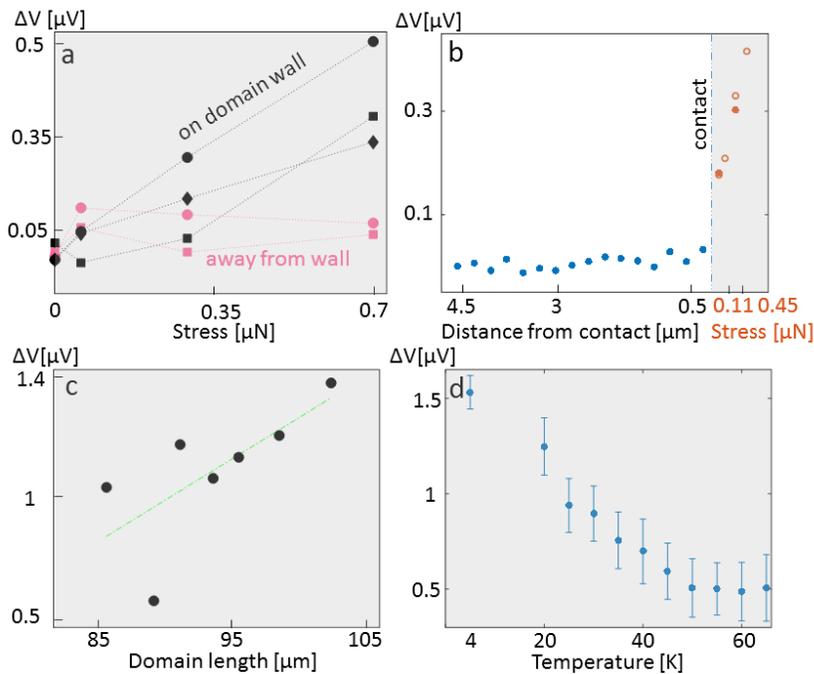

*Figure 2: $\Delta V$ increases with stress on domain walls and scales with domain size.* (a) $\Delta V$ as a function of stress for different locations. Black (magenta) symbols represent locations on (off) the domain wall, respectively. Lines are a guide to the eye. (b) Sample response as the tip approaches the sample along the Z direction. $\Delta V$ is plotted as a function of height (before contact, white background) and stress (after contact, grey background), does not change significantly as the tip approaches the sample. After contact we further push the piezo to apply increasing stress. The signal increases linearly with pressure. Empty circles indicate data taken from a different cooldown. (c) $\Delta V$ values (black circles) taken only from domains with defined borders (for size measurements) and the signal is taken as the strongest 10 pixels of each domain. The green line is a guide to the eye. $\Delta V$ increases with the size of the domain, supporting a non-local scenario. (d) Temperature dependence of $\Delta V$ modulations on a domain wall, showing an increase in strength below ~40 K (error bars represent the standard deviation from the mean $\Delta V$ modulation calculated over 10 scans for each temperature).

Microscopically the source of the $\Delta V$ signal can be ascribed to various mechanisms related to the local stress-induced structural changes. For example, local stress can change the number of oxygen vacancies[29] that are known to accumulate at domain walls[14]. The temperature evolution of the effects coincides with the observation of strong polarization inside twin walls below 40 K reported by Salje et al. [12,13] The coupling between the polarization inside the twin walls below 40 K and the conducting layer is akin to the usual electrostatic gating effect where an electric field attracts or repels electrons depending on its direction. In our case, polar domain walls act as local potential barriers, modulating the charge density and hence the local conductivity. The domain wall polarization is highly anisotropic and Fig. 1c shows that below 40 K the ratio between two perpendicular measuring directions decays to values below unity[20,26].

Domain wall polarization in STO is strongly stress dependent via the flexo-electric effect near the wall[24,29]. The stress dependent variations of the domain wall structure was previously observed in LAO with significant changes of the topology of the wall segments (Larkin lengths)[30]. We therefore relate our observed change in device resistivity in response to local pressure to a stress induced change in domain wall polarity and related changes of the domain wall topologies.

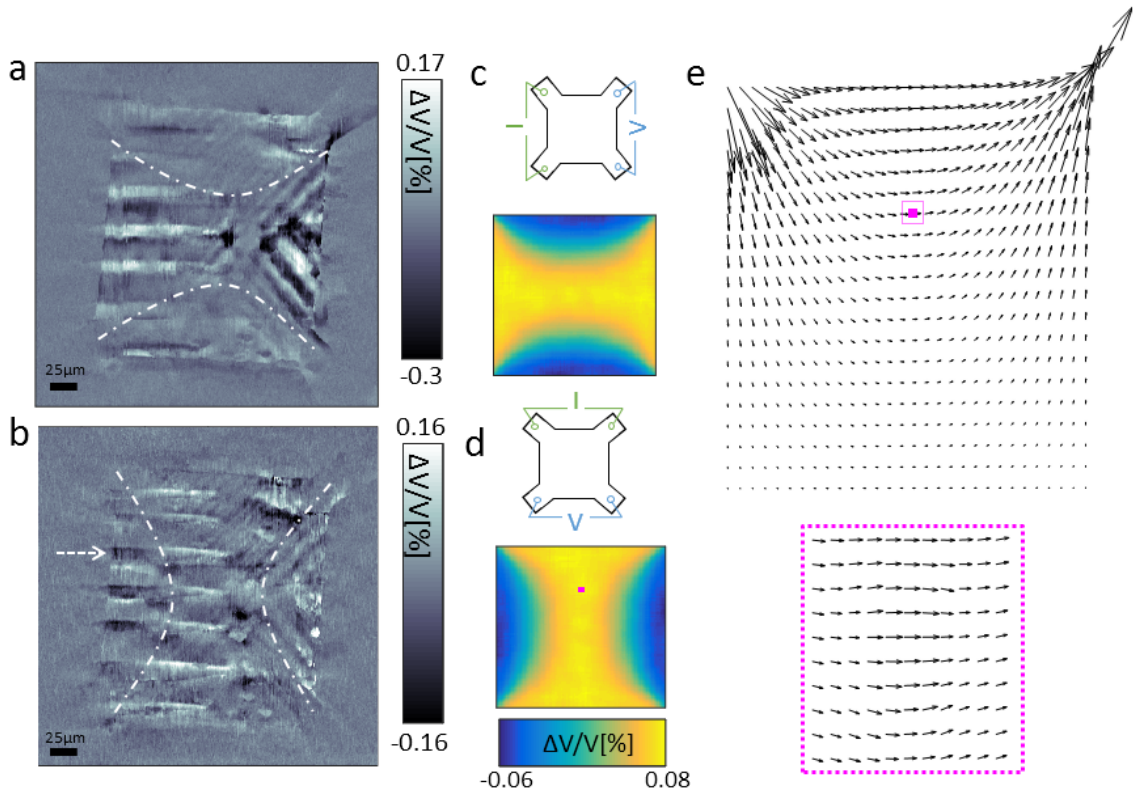

*Figure 3: Local change in resistivity diverts current flow, describes well the measured ΔV map, and suggests a non-local response.* (a-b) ΔV data, (a) ΔV when current flows along the left side and (b) along the top edge. (c-d) Calculated ΔV shows an hourglass shaped area of positive ΔV (yellow). For each pixel we calculated the difference in the four probe voltage as a result of increasing the resistivity at that pixel. (c) Horizontal hourglass shape observed when the current is injected / removed from the left corners and the four probe voltage is measured on the right side. (d) Vertical hourglass with current injected from the top corners. White dotted lines in a-b illustrate the hourglass shape on ΔV data. (e) Vector map. Calculation of current flow modulations in response to an increase in local resistivity at the pink square (details of calculations in Supplementary Information Fig. S6). Arrows represent size and direction of the current after the increase. The length of the single arrows at the corners represents the total current. The area zoomed at the dotted square demonstrates that the current path is slightly diverted to bypass the small more

*resistive region. The vector map explains one pixel in the hourglass shape (the pink dot in d). In the case of higher resistivity at the location of the pink dot more current is diverted towards the opposing voltage leads, resulting in a higher voltage reading. Thus, white colour inside the hourglass in a-b represents an increase in local resistivity.*

Considering the size of our contact point (area of physical contact between the tip and the sample, Methods) relative to the entire macroscopic sample, the measured $\Delta V$ values (Fig. 3a,b) are surprisingly high (0.16% $\Delta V/V$). Another intriguing feature is the sign switching half-way along the domain wall (see white arrow in Fig. 3b), and that the same wall does not switch sign in Fig. 3a. In order to understand these results we first examine the effect of a local change in resistivity on the current distribution in a homogeneous sample and calculate the expected four probe voltage.

In general, an increase in the resistivity of a small region in a sample alters the current flow. The local current is diverted to partially bypass that region (simulated in Fig. 3e). To estimate the expected $\Delta V$ we calculated the voltage change in response to local modulation of conductivity (detailed in Supplementary Information S6). The main feature apparent on the calculated $\Delta V$ map is the hourglass shape (horizontal in Fig. 3c and vertical in Fig. 3d). The reason for the hourglass shape is the way the current flows in the homogenous sample and the way it is diverted once we change the local conductivity (Fig. 3e). Increasing the resistivity in certain areas diverts the current towards the voltage leads, resulting in a higher voltage drop (positive $\Delta V$ signal), while increasing the resistivity outside the hourglass results in negative ΔV. The resultant hourglass shape is also apparent in the $\Delta V$ data (Figs. 3a-b, dashed line). The origin of the sign flip we observed in the $\Delta V$ signal along a single domain wall is now clear, as well as why it is observed only for the vertical current flow direction (Fig. 3b).

We note that although the simulation reproduced well the main features of the data (hourglass shape, sign switching and direction dependence), there is a considerable mismatch between the measured values and the values obtained by the simulation. The ΔV data is more than an order of magnitude larger than the expected signal from our calculations (Supplementary Information Fig S6). Interestingly, we found that we can only achieve values comparable to the data (maximum 0.2-0.3% on the domain walls) with a stress kernel that is much larger than the actual size of our contact area (Supplementary Information Fig. S6) – indicating that the stress kernel is larger than the contact area. The sharp termination of ΔV signal near the edge of the sample (Supplementary Information Fig. S7, contact diameter of ~ 0.2 µm) proves that this is not a simple decay of our stress dome. Further support comes from the ability to resolve dense domains (down to 2 µm spacing). We suggest that the response area extends along the domain wall. This scenario is supported by the increase of ΔV signal with domain length (Fig. 2c).

The similarity between the calculated ΔV to the measured ΔV map provides important insight into the underlying physical origin of the response. (a) The hourglass shape we observe in both directions of current injection (white dotted line Fig. 3a-b) indicates that we changed the local resistivity with stress. (b) The sign of ΔV signal is mainly white (positive ΔV change) inside the hourglass, indicating an increase in local resistivity as a result of local stress. (c) ΔV values calculated for reasonable contact areas are significantly smaller than the measured values. This suggests that the response to stress extends beyond our physical stress dome. (d) Finally, the most striking difference between the calculated and measured ΔV is the streaks of signal, indicating that the response to stress occurs on domain walls and not homogeneously over the sample.

Our results provide direct visualization of STO domain wall polarity, earlier deduced from resonant piezoelectric spectroscopy[12]. Salje *et al.* measured the mechanical vibrations induced by an a.c. voltage applied to an STO crystal. The magnitude of the response in STO, compared to the response detected in ferroelectric $BaTiO_3$ lead to the conclusion that polarity in STO resides in the ferroelastic domain walls. Our data shows that stress applied to certain regions in the sample affects the overall sample transport behaviour. The maps of the electric response to stress identify the domain walls as the "active spots". Given that in ferroelastic materials polarization is coupled to stress[30], our maps of stress response should represent the map of polarity. We find full correlation between the domain wall configuration and the map of polarity (Supplementary Information Fig. S4).

We show that local stress changes the wall polarity, depleting or accumulating carriers and thus affecting the local current flow. This is equivalent to altering or creating local electric fields. Electrostatic gating is a powerful tool for fundamental studies of complex oxide materials, providing nanoscale control of the electrostatic landscape to develop controllable devices[5,6,31]. Local gating by polar domain walls is similar; it is located near the conducting layer and it is only nanometres thick. A central advantage of domain walls is that they naturally occur near the interface. In addition, in clean STO, domain walls are highly mobile down to low temperatures[27]. In our samples we find no indication for significant domain wall pinning; their mobility is demonstrated by the substantial changes in domain wall configuration between cooldowns, and by their movement with electrostatic back gating[20–22,25]. An STO sample cooled below 105 K forms a network of domain walls that are sometimes well separated, but typically dense. In large samples the dense network blurs out the influence of the local electrostatic 'gating' by the walls. However, in small devices we can imagine using individual walls as local, well separated, gates that can be moved around by external electrostatic fields and tuned by stress. This opens the possibility of creating devices that are not fixed to a specific location and that can be in-situ created and tuned.

The current distribution at the LAO/STO interface was shown to modulate over STO domain structure. These modulations were observed by imaging the current flow with scanning SQUID[20]. Our current work suggests that the dominant mechanism for this modulation is domain wall polarization as suggested by Salje *et al*.[12] Other mechanisms are also possible, however: for example oxygen vacancies, which accumulate at the walls[14], could also donate free charges and increase local current flow near the wall. We note that previous studies at 4.2 K have imaged domain walls moving under back gate voltage [21,25] and that the location of the current modulations is also changed with back gate[20]. In the oxygen vacancies scenario, vacancies should move at 4.2K with the wall.

We observed a change in the local resistivity at the wall in response to local stress. The explanation we propose is that stress applied by our tip changes the wall polarity, in magnitude and/or direction. In this manner the effect of the wall polarity on the nearby conducting layer also changes. This small change in the local current flow in the conducting layer is read by our system. These stress induced changes in the wall polarity serve as local electrostatic gates and locally tune the conducting layer near the wall. We emphasise, that below 40 K, the walls become polar and locally modulate the current flow even without the application of stress[12,20]. Our findings support this scenario, as we only detect response to stress below 40 K. The local stress only tunes the polarization as the walls were already polar. This is somewhat similar to an explanation proposed by Stolichnov *et al.* for conduction in BFO ferroelectric domain walls, where stress tuned polarization affects the local electronic states.[32] Another possible link between stress and wall polarity is stress related change of the recently discovered vortex motion inside the domain walls[13]. According to simulations by Zykova-Timan *et al*.[13], the polar moments inside the walls form vortex structures that are expected to be unpinned and highly mobile. These vortices should move under external stimuli such as stress. Their movement would also cause a change in polarization, further tuning the local current flow, which was already modulated by the initial polarization of the wall. Another possible mechanism that relates stress to local change in resistivity is movement of the domain wall due to stress. In clean STO domain walls are still mobile at 4.2 K under external stimuli such as back gate voltage[20,21]. Applying stress near the walls may result in their motion. Motion of polarized domain walls can divert some of the current flow and impact the device's resistance.

Lastly, by mapping the voltage response to the applied stress we identified that the local resistivity predominantly increases with applied stress. As we apply pressure to the wall we increase the total polarity of the wall either by aligning more polarization moments to the same direction or by increasing the size of the moments. Subsequently, more electrons from the 2DEG are then needed in order to screen this extra polarization. These screening electrons are now localized and do not contribute to the conductance, therefore the resistance rises.

In conclusion, our data provides direct visualization of the polarity inside the STO domain boundaries and show how they act as local gates and tune currents in nearby conducting layers. We find an unexpectedly large electrical response to local stress in LAO/STO devices. The response peaks on stripe-like features, which we identify as STO ferroelastic domain walls. The response is stronger than expected for the physical contact area, indicating an extended response along the domain wall. STO domain walls are mobile and are at the nanoscale. The ability to control the electrostatic landscape with these walls provides the ground for future in-situ fabrication of electronic and superconducting circuits. Further, the LAO/STO interface was originally believed to be a 2D system but apparently is filled with quasi-1D channels. Our work provides direct imaging of the polarity at the STO domain walls and a direct link between the channel flow and this polarity. Finally we wish to draw the attention of the reader to recent results on wall motion near quantum critical points (QCP).[33] Under such conditions enhanced quantum fluctuations change the nature of the domain-wall kinetics from thermally activated motion to temperature-independent tunneling motion. As STO at 4.2K is close to a QCP [9] we may expect wall tunneling motion.

# Methods:

Five unit cells of LAO film were grown on top of $TiO_2$ terminated (001) STO substrate. The LAO/STO was patterned to a square (200 µm) for van der Pauw measurements[34], (illustration in Fig. 1a). We chose a square sample geometry rather than the conventional Hall bar because in this way small changes in the current flow can be easily monitored. In a square geometry most of the current flows directly between the current leads; the amount of current that reaches the voltage leads is exponentially small. Therefore, even small changes in the current flow result in a significant change in the recorded voltage, about 60 times more than in a Hall bar geometry[35].

Using the van der Pauw configuration and the tip of our scanning SQUID chip we constructed a scanning stress microscope that is extremely sensitive to small changes in current flow and that can be read as a change in four probe voltage. We used the scanning stress microscope to detect small changes in current as a function of location of the applied stress. The force range we used with the estimated contact area reach stress of up to $10^8$ Pascal, well within the elastic regime. The response we recorded was completely reversible and reproducible. However, we also recorded an irreversible component that appeared as line noise in our scans (see Supplementary Information Fig. S5). Using smaller tips and stronger forces it should be possible to reach the irreversible regime[29].

For calculating the contact area (a) we used the Hertzian contact formula:

$$Contact\ radius\ a^3 = \frac{3F}{8} \frac{\frac{(1-v_1)}{E_1} + \frac{(1-v_2)}{E_2}}{1/d_1}$$

Where *F* is the applied force, v, E, and *d* are the Poisson ratio, Young's modulus and radius of curvature. The subscripts refer to: "1" for silicon tip silicon[36] and "2" for the STO substrate[37,38].

**Data availability.** The authors declare that the data supporting the findings of this study are available within the paper and its supplementary information files.

# Acknowledgements

Y.F., N.H., Y.S. and B.K. were supported by the European Research Council grant ERC-2014-STG-639792 and Israel Science Foundation grant ISF-1102/13. Z.C., Y.H. and H.Y.H. acknowledge support by the Department of Energy, Office of Basic Energy Sciences, Division of Materials Sciences and Engineering, under contract DE-AC02–76SF00515 (Y.H.), and the Gordon and Betty Moore Foundation's EPiQS Initiative through Grant GBMF4415 (Z.C.).

# Author contributions

Y.F., Y.S. and B.K. designed the experiment and performed the measurements. C.B., Y.X., Z.C., Y.H. and H.Y.H. provided the samples, N.H. contributed to the simulations and E.K.H.S. contributed to the interpretation of the data. Y.F. and B.K. prepared the manuscript with input from all co-authors.

# Notes

The authors declare no competing financial interests.